\documentstyle[11pt]{article}

\title{\bf Pomeron with a running coupling constant:\\
intercept and slope}
\author{M. A. Braun \thanks{On leave of absence from the Department of
high-energy physics,
 University of S. Petersburg, 198904 S. Petersburg, Russia}  \\
Department of Particle Physics, University of Santiago de Compostela,\\
15706 Santiago de Compostela, Spain}
  \date{October 1994}
\def\beq{\begin{equation}}
\def\eeq{\end{equation}}
\def\noi{\noindent}
\def\oq{\omega(q)}

\def\eq{\eta (q)}
\def\ea{\eta (q_{1})}
\def\eb{\eta (q_{2})}
\def\ec{\eta (q'_{1})}
\def\ed{\eta (q'_{2})}
\begin{document}
\maketitle
\medskip
\noi{\bf Abstract}
The equation for two reggeized gluons with a  running QCD coupling  constant,
proposed earlier on
the basis of the bootstrap condition, is investigated by the variational
technique in the vacuum channel. The pomeron intercept and slope are
calculated as a function of the infrared cutoff parameter $m$ ("the gluon
mass"). The intercept depends weakly on $m$ staying in the region $0.3--0.5$
for  values  of $m$ in the range $0.3--1.0\ GeV$. The slope goes down from
$5.0\ GeV^{-2}$ to $0.08\ GeV^{-2}$ in this range. The optimal value of $m$
seems to lie in the region $0.6--0.8\ GeV$ with the intercept $0.35$ and slope
$0.35--0.15$. The character of the pomeron singularity: pole or cut, is also
discussed.
\vspace{3 cm}

{\Large\bf US-FT/17-94}
\newpage

{\bf 1. Introduction.}
We have recently proposed a new method to incorporate a running coupling
constant into the equation governing the behaviour of two reggeized gluons
[ 1 ]. It is based on  the so-called "bootstrap condition" of L.Lipatov [ 2
], which requires that the solution in the gluon colour channel should be
the reggeized gluon itself. The bootstrap condition leads to a relation
between the gluon interaction $U$ and its Regge trajectory $\omega$: both
have to be expressed via the same function $\eta$, so that in the momentum
space
\beq
\oq=-(3/2)\eq\int (d^{2}q_{1}/(2\pi)^{2})/\ea\eb
\eeq
and
\beq
U(q,q_{1},q'_{1})=-3(\ea/\ec+\eb/\ed)/\eta(q_{1}-q'_{1})+3\eq/\ec\ed
\eeq
In (2) $q$ is the total momentum of the two interacting gluons. In both
equations $q_{1}+q_{2}=q$.

The asymptotical behaviour of $\eq$ at high momenta can be found from the
comparison of the pomeron equation at $q=0$ and the standard
Gribov-Lipatov-Altarelli-Parisi evolution equation in the double
logarithmic approximation. With a running QCD coupling constant $g(q)$ we
obtain at high $q$
\beq
1/\eq=g^{2}(q)/2\pi q^{2}=2\pi/bq^{2}\ln q^{2}/\Lambda^{2}
\eeq
where
\beq
b=(11-(2/3)N_{F})/4
\eeq
and $\Lambda\simeq 200\ Mev$ is the standard QCD parameter.
With a fixed coupling constant (the BFKL pomeron [ 3 ]) $1/\eq$ results
proprtional to $q^{2}$. It follows that the introduction of a running
coupling constant changes the pomeron equation already at leading order, so
that it cannot be studied by a perturbative approach like in [ 4 ].

The behaviour of $\eq$ at low $q^{2}$ remains unknown and, as we want to
stress, hardly amenable to any theoretical approach, since it belongs to the
nonperturbative confinement region. More than that, the two-gluon equation
itself looses sense at low $q\sim\Lambda$, where gluons cease to exist as
individual particles. Therefore the behaviour of $\eq$ at low $q^{2}$
 plays the role of a boundary condition to be supplemented to the pomeron
equation, which effectively takes into account the confinement effects.
One can at most try to parametrize it in some simple manner and study the
consequences. In [ 1 ] we have chosen a parametrization
\beq
1/\eq=2\pi/b(q^{2}+m^{2})\ln ((q^{2}+m^{2})/\Lambda^{2})
\eeq
where the parameter $m\geq\Lambda$ has a meaning of an effective gluon mass.

The asymptotical behaviour of both the gluon trajectory and interaction with
the choice (5) was studied in [ 1 ].  Taking for them simple
analytic expressions which interpolate between low and high values of their
arguments, the pomeron intercept was also studied
 by the variational technique in [ 1 ].

In this note we investigate the pomeron intercept with exact values for the
trajectory and interaction, which follow from (1) and (2) with the choice
(5). We confirm that the intercept depends very weakly on the value
of the gluon mass $m$, as was found in [ 1 ] for approximate $\omega$ and
$U$. Its absolute values turn out, however, somewhat higher: for $m$ growing
from $0.3\ GeV$ to $2\ GeV$ the intercept falls from $0.5$ to $0.25$.

Since we know the pomeron equation with a running coupling constant also for
$q\neq 0$, we are in a position to calculate the slope of the pomeron
trajectory at $q=0$. It happily results positive, its values depending very
strongly on $m$: in the same interval $m=0.3--2.\ GeV$ the slope goes down
from $5.0\ GeV^{-2}$ to $0.01 \ GeV^{-2}$. From the experimental observations
then the optimal value of $m$ lies in the region of the $\rho$-meson mass
$m=0.6--0.8\ GeV$.

We finally discuss the character of the pomeron singularity in the $j$-plane.
As it seems, it depends on the chosen behaviour of $\eq$ at $q\rightarrow 0$.
For the choice (5) it turns out to be a moving pole, although we cannot
exclude that other choices may result in a moving cut.

{\bf 2. The intercept.}
For the forward scattering, $q=0$, the pomeron equation aquires the form of
a Shroedinger equation in two dimensions
\beq
H\psi =E\psi
\eeq
where the Hamiltonian is a sum
\beq H=T+U+Q\eeq
of a kinetic energy
\beq
T(q)=-2\oq
\eeq
a local interaction $U(r)$ with a Fourier transform
\beq
U(q)=-6/\eq
\eeq
and a separable interaction
\beq
Q=(1/12)\eta(0)|U\rangle\langle U|
\eeq
The intercept $\Delta$ is the ground state energy of (6) with a minus sign.

In [ 1 ] the analytic forms were used for $T(q)$ and $U(r)$ which
interpolate between small and large values of their respective arguments
\beq
T(q)=T(0)+\ln\ln(q^{2}+m^{2}),\ \ U(r)=\ln\ln(1/r^{2}+m^{2})
\eeq
(in units $3/b$, $\Lambda=1$). Numerical calculations with $\eq$ given by (5)
lead to $T(q)$ and $U(r)$ shown in Table 1. for $m=3.0$ in the regions
where they noticeably differ from their approximate expressions (11) (shown by
dashed curves). As one observes, the largest error in using (11) is for $U(r)$
at large $r$, where the exact potential falls exponentially, whereas an
approximate one only as $1/r^{2}$. Since the potential is attractive, one
might think that with the approximate potential the ground state level should
lie deeper. However the approximate kinetic energy lies systematically above
the exact one. This effect turns out to be dominating.

Taking exact numerical values for $T(q)$ and $U(r)$ we investigated the
ground state energy $E_{0}$ by the standard variational method. As in [ 1 ]
trial functions were chosen
 as linear combinations of the
two-dimensional harmonic oscillator functions for zero angular momentum
\beq
\psi(r^{2})=\sum_{k=0}^{N} c_{k}{\mbox L}_{k}(ar^{2})\exp (-ar^{2}/2)
\eeq
with ${\mbox L}_{k}$ the Lagerre polinomials and $a$ and $c_{k}$
variational parameters. Up to 16  polinomials have been included
in the calculations. The resulting intercept is shown in Table 2. (the
second column) as a function of $m$ for $\Lambda=0.2\ GeV$ and three flavours
($b=9/4$). As one observes, it depends on $m$ rather weakly except for the
values of $m$ quite close to $\Lambda$. For reasonable values of $m$ in the
range $0.3--1.\ GeV$ it goes down from $0.5$ to $0.3$. Comparing these values
with the ones found in [ 1 ] with (11) we find that the exact trajectory and
interaction lead to the intercept twice as high.

{\bf 3. The slope.}
To determine the slope at $q=0$ we can use the perturbative approach.
For small values of $q$ we present the Hamiltonian in the form
\beq
H=H_{0}+W(q)
\eeq
where $H_{0}$ is the Hamiltonian for $q=0$ and $W$ is the perturbation
proportional to $q^{2}$. The slope $\alpha'$ is then calculated as the
average of $W$ in the ground state $\psi_{0}$ at $q=0$ divided by $q^{2}$.

At $q\neq 0$ the pomeron equation retains its form (6). The components of
the Hamiltonian (7) are now given by
\beq
T=-\omega_{1}-\omega_{2}
\eeq
\beq
U=-3
(\sqrt{\eta_{1}/\eta_{2}}V\sqrt{\eta_{2}/\eta_{1}}+
\sqrt{\eta_{2}/\eta_{1}}V\sqrt{\eta_{1}/\eta_{2}})
\eeq
\beq
Q=3\eq |1/\sqrt{\eta_{1}\eta_{2}}\rangle\langle
1/\sqrt{\eta'_{1}\eta'_{2}}|
\eeq
Here and in the following
$\omega_{1}=\omega (q_{1}),\  \eta_{1}=\eta (q_{1})$ and so on.

 Let $q_{1}=(1/2)q+l,\ q_{2}=(1/2)q-l$. To find $W(q)$ we develop
 (14)-(16) in powers of $q$ up to the second order. Straightforward
calculations lead to
\beq
T=T_{0}-(1/8)[qr[qr,T_{0}]]
\eeq
\beq
U=U_{0}+(1/2)[(ql)\zeta_{1}[(ql)\zeta_{1},U_{0}]]
\eeq
\beq
Q=Q_{0}+q^{2}\zeta_{3}Q_{0}-
\{(1/4)q^{2}\zeta_{1}+(1/2)(ql)^{2}(\zeta_{2}-\zeta_{1}^{2}),Q_{0}\}
\eeq
where $T_{0}, U_{0}$ and $Q_{0}$ correspond to $q=0$ and are given by
(8)-(10). We have denoted
\[\zeta_{1}(l)=(\ln\mu+1)/\mu\ln\mu,\ \ \zeta_{2}(l)=1/\mu^{2}\ln\mu,\ \
\zeta_{3}=\zeta_{1}(0)\]
where $\mu=m^{2}+l^{2}$. Note that in the double commutators (17) and (18) one
can change $T_{0}$ and $U_{0}$ to $T_{0}+U_{0}=H_{0}-Q_{0}$, since the added
terms commute with the respective operators.
In this manner we finally
obtain
\[
W=-(1/8)[qr[qr,H_{0}-Q_{0}]]
+(1/2)[ql\zeta_{1}[ql\zeta_{1},H_{0}-Q_{0}]]\]
\beq +q^{2}\zeta_{3}Q_{0}-
\{(1/4)q^{2}\zeta_{1}+(1/2)(ql)^{2}(\zeta_{2}-\zeta_{1}^{2}),Q_{0}\}
\eeq

Taking the average of (20) in the ground state $\psi_{0}$  we present the
corresponding change in energy $\Delta E$ as a sum of three terms
\beq
\Delta E=\sum_{n=1}^{3}\Delta E^{(n)}
\eeq
Here terms with $n=1,2$ and $3$ correspond to parts of (20) proportional to
the ground state energy
$E_{0}$, containing $T_{0}+U_{0}$ and containing $Q_{0}$ respectively.
\beq
\Delta E^{(1)}=E_{0}\langle (-1/4)(qr)^{2}+(ql)^{2}\zeta_{1}^{2}\rangle
\eeq
\beq
\Delta E^{(2)}=\langle(1/4)(qr)(T_{0}+U_{0})(qr)-(ql)\zeta_{1}(T_{0}+U_{0})
(ql)\zeta_{1}\rangle\eeq
and
\beq
\Delta E^{(3)}=3\eta_{0}\langle\psi_{0}|((1/4)(qr)^{2}
+q^{2}(\zeta_{3}-(1/2)\zeta_{1})-(ql)^{2}\zeta_{2})V\rangle\langle
V|\psi_{0}\rangle
\eeq
where we denoted $\langle\psi_{0}|A|\psi_{0}\rangle\equiv\langle A\rangle$.

In most of these averages one can immediately integrate over the azimuthal
angle substituting
$(ql)^{2}\rightarrow (1/2)q^{2}l^{2}$ and
 $(qr)^{2}\rightarrow (1/2)q^{2}r^{2}$.
 Two averages, however, are a bit more complicated.
\[ A_{1}\equiv\langle(qr)T_{0}(qr)\rangle,
\ \ A_{2}\equiv\langle(ql)
\zeta_{1}U_{0}(ql)\zeta_{1}\rangle\]
The
first one is evidently a momentum space average in the state
$(qr)\psi_{0}(l^{2})=2i(ql)\psi'_{0}(l^{2})$, so that
$ A_{1}=2q^{2}\langle\psi'_{0}|l^{2}T_{0}|\psi'_{0}\rangle$.
The second average, also in the momentum space,
reduces to that of a nonlocal potential $U_{1}$:
$A_{2}=(1/2)q^{2}\langle U_{1}\rangle$
where  $U_{1}$ possesses a kernel
\beq
U_{1}(l_{1},l_{2})=-6\zeta_{1}(l_{1})\zeta_{1}(l_{2})
(l_{1}l_{2})/\eta(l_{1}-l_{2})
\eeq

The final formula for the
slope consists of three terms in correspondence with (22)-(24)
\beq
\alpha'=\sum_{n=1}^{3}\alpha'_{n}
\eeq
They  are
\beq
\alpha'_{1}=E_{0}\langle (-1/8)r^{2}+(1/2)l^{2}\zeta_{1}^{2}\rangle
\eeq
\beq
\alpha'_{2}=\langle(1/8)r^{2}U_{0}-(1/2)l^{2}\zeta_{1}^{2}T_{0}
-(1/2)U_{1}\rangle+(1/2)\langle\psi'_{0}|l^{2}T_{0}|\psi'_{0}\rangle
\eeq
\beq
\alpha'_{3}=3\eta_{0}\langle\psi_{0}|((1/8)r^{2}
+\zeta_{3}-(1/2)\zeta_{1}-(1/2)l^{2}\zeta_{2})V\rangle\langle
V|\psi_{0}\rangle
\eeq

We have calculated $\alpha'$ according to these formulas taking for the
ground state $\psi_{0}$ the trial function (12) with the parameters which
minimize the energy. The results are shown in Table 2. (the third column).
They strongly depend on  $m$, reaching quite large values for $m$ close to
$\Lambda$ and rapidly falling with its growth to values of the order $0.01\
GeV^{-2}$ for $m=10\lambda$.

{\bf 4. Pole or cut?}
We finally want to comment on the character of the pomeronic singularity in
the $j$ plane for our eqation. It is easy to show that with the choice of
$\eq$ according to (5) it is a moving pole. To prove it we have to
demonstrate that the ground state wave function $\psi_{0}$ is normalizable.

With $\eq$ given by (5), the interaction falls exponentially for large
intergluon distances $r\rightarrow\infty$,
  so that the equation reduces to
\beq
(T(q)-E)\psi(r)=0\eeq
where $q^{2}=-\Delta$. Evidently the asymptotics of the solution is a linear
combination of  zero order Bessel functions
\beq
\psi(r)\simeq a{\mbox H}^{(1)}_{0}(qr)+b{\mbox H}^{(2)}_{0}(qr)
\eeq
where $q$ is determined from the equation
\beq
T(q)=E
\eeq
 However, for positive $q^{2}$ the kinetic energy $T(q)$ is
greater than a positive threshold value $T(0)$ (Table. 1). Then for $E\geq
T(0)$
the asymptotics (31) is oscillating and $\psi$ is not normalizable. For
$E<T(0)$, and for $E<0$ in particular,  Eq. (32) gives pairs of complex
conjugate solutions
$q=q_{1}\pm q_{2}$ with $q_{2}>0$. Then evidently the function $\psi(r)$
becomes exponentially falling at large $r$ like $\exp(-q_{2}r)$

To show its normilizability we have additionally to study its behaviour at
$r=0$, that is, at large momenta $q$. With the asymptotics of the kinetic
energy clear from (11), the equation at $q\rightarrow\infty$ becomes
\beq
\ln\ln q^{2}\psi(q)+bm^{2}\ln m^{2}/q^{2}\ln q^{2}=
(1/\pi)\int d^{2}q'\psi(q')/((q-q')^{2}+m^{2})\ln ((q-q')^{2}+m^{2})
\eeq
Two alternatives are possible: either
\beq\psi(r=0)=\int (d^{2}q/2\pi)\psi(q)<\infty\eeq
or $\psi(r=0)=\infty$. We  want to show that the first one is realized. In
fact, assume that
 $\psi(r=0)=\infty$. Introduce a function $f(\xi)=q^{2}\ln q^{2}\psi(q)$ with
 $\xi=\ln\ln q^{2}$. By assumption
 \[\int d\xi\,f(\xi)=\infty\]
 After calculating the asymptotical behaviour of the integral term in (33) one
arrives at an equation for $f$ of the form
\[\xi f(\xi)=\int^{\xi}d\xi'(f(\xi')+f(\xi))\]
which can never be satisfied, since the second term on the righthand side
exactly cancels the lefthand side term. Thus (34) is true. Then the integral
term in (33) behaves like $1/q^{2}\ln q^{2}$ at high $q$, which contribution
should be cancelled by the one from the separable interaction (the second
term  on the lefthand side). The first term, coming from the kinetic energy,
then represents a subdominant correction. With (34) fulfilled, the function
$\psi(r)$ is well-behaved for all $r$ and is normalizable. Then solutions for
$E_{0}<0$ should belong to a discrete spectrum.

We have to point out, however, that this conclusion depends heavily on our
choice of $\eq$, which assumes that it is different from zero at $q=0$. As
discussed in the beginning, actually we do not know anything about the
behaviour of $\eq$ at $q<<\Lambda$ nor does it make any physical sense.
More general parametrizations,
like
\[\eq=(b/2\pi)(q^{2}+m_{1}^{2})\ln (q^{2}+m^{2})\]
with two different masses $m$ and $m_{1}$ are equally admissible. In
particular we can take $m_{1}=0$ and choose
\beq
\eq=(b/2\pi)q^{2}\ln (q^{2}+m^{2})
\eeq
when $\eta(0)=0$. This choice seems to have some advantages: now the
trajectory will go through 1 at $q=0$ ( for the "physical" gluon) and also
the solution will satisfy the condition $\psi(q=0)=0$, which follows from
the colour current conservation if one goes down to $q=0$. However these
advantages are completely spurious: with confinement one can never go as far
as $q=0$  and should  always stay at momenta $q>\Lambda$ in coloured channels.
So, in our opinion, the choice (35) is nothing better than our earlier (5).

It however leads to a considerably more complicated gluonic Hamiltonian.
With (35) the kinetic energy and interaction do not exist separately, but have
to be taken together in the integral kernel exactly in the same manner as
with the BFKL equation [ 3 ]. The infrared behavior of this new Hamiltonian
is completely different from the one with (5) and becomes  similar to
that of the BFKL Hamiltonian. With that, the equation with a running
coupling constant does not possess the scaling symmetry and hardly admits an
analytic study. We cannot say anything definite about the pomeron singularity
for it: it might well be a  cut as for the BFKL equation (only a moving cut,
contrary to the BFKL case).

Thus we conclude that in the real world with a running coupling constant the
equation for the pomeron is sensitive to the confinement effects, that is, to
the behaviour of the trajectory and interaction in the infrared region. The
dependence on this behaviour is minimal (and quite weak, indeed) for the
intercept, whose values, after all, result not very different from those
extracted from the BFKL pomeron for reasonable values of the coupling
constant. However the infrared behavior has a strong influence on the slope
value and may be essential for the character of the pomeron singularity.
\vspace{1 cm}

{\bf Acknowledgments.}
The author is deeply grateful to E.Levin, L.Lipatov,
 C.Pajares and N.Armesto for
fruitful and constructive discussions. He also expresses his gratitude
to the General Direction of the Scientific and Technical Investigation
(DGICYT) of Spain for financial support.
\vspace{1 cm}

{\bf References.}

\noi 1. M.Braun, Univ. of Santiago preprint US-FT/11-94\\
\noi 2. L.Lipatov in "Petrturbative QCD", Ed.  A.Mueller, Advanced
Series in High Energy Physics (World Scientific, Singapore, 1989)\\
\noi 3. E.Kuraev, L.Lipatov and V.Fadin, Sov. Phys. JETP, {\bf 45}
(1977) 199\\
Ya.Balitzky and L.Lipatov, Sov. Phys. J.Nucl. Phys.,{\bf 28} (1978)822\\
\noi 4. V.Fadin and L.Lipatov, Nucl. Phys. {\bf B406} (1993) 259

\newpage \vspace*{3 cm}
{\Large\bf Table 1. Exact and approximate gluon trajectory and potential}
\vspace {1 cm}

\begin{tabular}{|r|r|r||r|r|r|}\hline
$q^{2}$&$T$&$T_{approx}$& $r^{2}$&$-U$&$-U_{approx}$\\\hline
0.37 & 0.133 & 0.138 & 0.0183& 1.319  & 1.293  \\\hline
1.00 & 0.147 & 0.178 & 0.0498& 1.044  & 1.008  \\\hline
2.72 & 0.241 & 0.372 & 0.135 &  0.695 & 0.637  \\\hline
7.39 & 0.593 & 0.767 & 0.368 &  0.302 & 0.241  \\\hline
20.1 & 1.013 & 1.139 & 1.00  & 0.0426 &0.0468  \\\hline
54.6 & 1.293 & 1.424 & 2.72  & 2.78E-4& 6.77E-3\\\hline
148. & 1.515 & 1.646 &  7.39 &4.03E-10& 9.25E-4\\\hline
403. & 1.698 & 1.829 &       &        &        \\\hline
1097 & 1.852 & 1.983 &       &        &        \\\hline
2981 & 1.985 & 2.116 &       &        &        \\\hline
8103 & 2.103 & 2.234 &       &        &        \\\hline
\end{tabular}
\vspace{1 cm}

{\Large\bf Table 1. captions}
\vspace{1 cm}

The first column gives  values of the gluon momentum squared in units
$\Lambda^{2}$. In the second and third columns the corresponding exact
and approximate (Eq. (11)) values
 of the gluon trajectory are
presented in units $6/b$. The fourth column gives the
 values of the intergluon distance squared
in units $\Lambda^{-2}$. In the last two columns the corresponding
exact and approximate values of the gluon potential (with minus sign) are
given (in units $3/b$)

\newpage \vspace*{3 cm}
{\Large\bf Table. Pomeron intercept and slope}
\vspace {1 cm}

\begin{tabular}{|r|r|r|}\hline
$m$&$\Delta $& $\alpha'$\\\hline
0.210&   1.02&   86.0 \\\hline
0.245&  0.687&   13.8\\\hline
0.283&  0.579&   5.77\\\hline
0.4  &  0.451&   1.32\\\hline
0.6  &  0.372&  0.346\\\hline
0.8  &  0.336&  0.150\\\hline
1.0  &  0.311& 0.0812\\\hline
1.41 &  0.280& 0.0333\\\hline
2.0  &  0.256& 0.0138\\\hline
\end{tabular}
\vspace{1 cm}

{\Large\bf Table 2. captions}
\vspace{1 cm}

The first column gives values of the gluon mass  in $GeV$
In the second column  the corresponding pomeron intercepts are
presented. The third column gives the pomeron slopes in $GeV^{-2}$.
The QCD parameter $\Lambda$ has been taken to be $0.2\ GeV$ for three flavours
($b=9/4$).

 \end{document}